\documentclass[fleqn,10pt, tikz,border=10pt]{wlscirep}
\usepackage{amsmath}
\usepackage{graphicx}  
\usepackage{nameref}
\usepackage{hyperref}
\usepackage{setspace}

\usepackage{caption}

\usepackage{subcaption}  
\usepackage[utf8]{inputenc}
\usepackage{tabularx,longtable,multirow,makecell}
\usepackage[figurename=Fig.]{caption}
\usepackage[nolist]{acronym}

\usepackage{multibib}
\newcites{methods}{Methods References}

\title{Optical Computation-in-Communication enables low-latency, high-fidelity perception in telesurgery}

\author[1,6,*]{Rui Yang}
\author[1,6]{Jiaming Hu}
\author[2,3,*]{Jian-Qing Zheng}
\author[1]{Yue-Zhen Lu}
\author[2]{Jian-Wei Cui}
\author[1]{Qun Ren}
\author[1]{Yi-Jie Yu}
\author[1]{John Edward Wu}
\author[4]{Zhao-Yu Wang}
\author[1]{Xiao-Li Lin}
\author[5]{Dandan Zhang}
\author[1]{Mingchu Tang}
\author[1]{Christos Masouros}
\author[1]{Huiyun Liu}
\author[1,*]{Chin-Pang Liu}

\affil[1]{Department of Electronic and Electrical Engineering, University College London, London WC1E 7JE, UK}
\affil[2]{CAMS Oxford Institute, Nuffield Department of Medicine, University of Oxford, Oxford OX3 7BN, UK}
\affil[3]{The Kennedy Institute of Rheumatology, University of Oxford, Oxford OX3 7FY, UK}
\affil[4]{Department of Computing, Imperial College London, London SW7 2AZ, UK}
\affil[5]{Department of Bioengineering, Imperial College London SW7 2AZ, UK}
\affil[6]{Rui Yang, Jiaming Hu contributed equally}
\affil[*]{Corresponding authors: rui.yang.18@ucl.ac.uk; jianqing.zheng@ndm.ox.ac.uk; chin.liu@ucl.ac.uk}
\definecolor{high}{RGB}{255,144,30}
\definecolor{Jianqing}{RGB}{0,155,155}
\definecolor{rui}{RGB}{30,30,255}
\definecolor{Jiaming}{RGB}{110 139 61}
\begin{abstract}
Artificial intelligence (AI) holds significant promise for enhancing intraoperative perception and decision-making in telesurgery, where physical separation impairs sensory feedback and control. Despite advances in medical AI \cite{Varghese2024,Esteva2017,Singhal2023} and surgical robotics \cite{Ciuti2025,Sheetz2020,Marcus2024,Shademan2016}, conventional electronic AI architectures remain fundamentally constrained by the compounded latency from serial processing of inference and communication. This limitation is especially critical in latency-sensitive procedures such as endovascular interventions, where delays over 200 ms can compromise real-time AI reliability and patient safety \cite{Motiwala2025,Larcher2023,barba2022remote}. Here, we introduce an Optical Computation-in-Communication (OCiC) framework that reduces end-to-end latency significantly by performing AI inference concurrently with optical communication. OCiC integrates Optical Remote Computing Units (ORCUs) directly into the optical communication pathway, with each ORCU experimentally achieving up to 69 tera-operations per second per channel through spectrally efficient two-dimensional photonic convolution. The system maintains ultrahigh inference fidelity within 0.1\% of CPU/GPU baselines on classification and coronary angiography segmentation, while intrinsically mitigating cumulative error propagation, a longstanding barrier to deep optical network scalability \cite{Fu2024}. We validated the robustness of OCiC through outdoor dark fibre deployments, 
confirming consistent and stable performance across varying environmental conditions. 
When scaled globally, OCiC transforms long-haul fibre infrastructure into a distributed photonic AI fabric with exascale potential, enabling reliable, low-latency telesurgery across distances up to 10,000 km and opening a new optical frontier for
distributed medical intelligence.

\end{abstract}

\begin{document}

\flushbottom
\maketitle

\thispagestyle{empty}


\acrodef{AI}[AI]{Artificial Intelligence}
\acrodef{SMF}[SMF]{Single-Mode Fibre}
\acrodef{PCI}[PCI]{Percutaneous Coronary Intervention}
\acrodef{ASD}[ASD]{Average Surface Distance}
\acrodef{CNN}[CNN]{Convolution Neural Network}
\acrodef{CPU}[CPU]{Central Processing Unit}
\acrodef{ECDF}[ECDF]{Empirical Cumulative Distribution Function}
\acrodef{GPU}[GPU]{Graphics Processing Unit}
\acrodef{HD}[HD]{Hausdorff Distance}
\acrodef{DSC}[DSC]{Dice Similarity Coefficient}
\acrodef{NDFF}[NDFF]{National Dark Fibre Facility}
\acrodef{OCiC}[OCiC]{Optical Computation-in-Communication}
\acrodef{ORCU}[ORCU]{Optical Remote Computing Unit}
\acrodef{MNIST}[MNIST]{Modified National Institute of Standards and Technology}
\acrodef{2D}[2D]{two-dimensional}
\acrodef{TOPS}[TOPS]{tera-operations per second}
\acrodef{EO}[EO]{Electro-Optic}
\acrodef{MZM}[MZM]{Mach-Zehnder modulator}
\acrodef{OSSM}[OSSM]{Optical Spectral Shaping Module}
\acrodef{DSP}[DSP]{Digital Signal Processing}
\acrodef{EDFA}[EDFA]{Erbium-Doped Fibre Amplifier}
\acrodef{ROC}[ROC]{receiver operating characteristic}
\acrodef{AUROC}[AUROC]{area under the ROC curve}
\acrodef{AWGN}[AWGN]{additive white Gaussian noise}
\acrodef{CW}[CW]{continuous-wave}

\acrodef{LMICs}[LMICs]{Low- and middle-income countries}

\section*{Main}

\begin{figure*}[!ht]
    \centering 
    \includegraphics[width=1\linewidth]{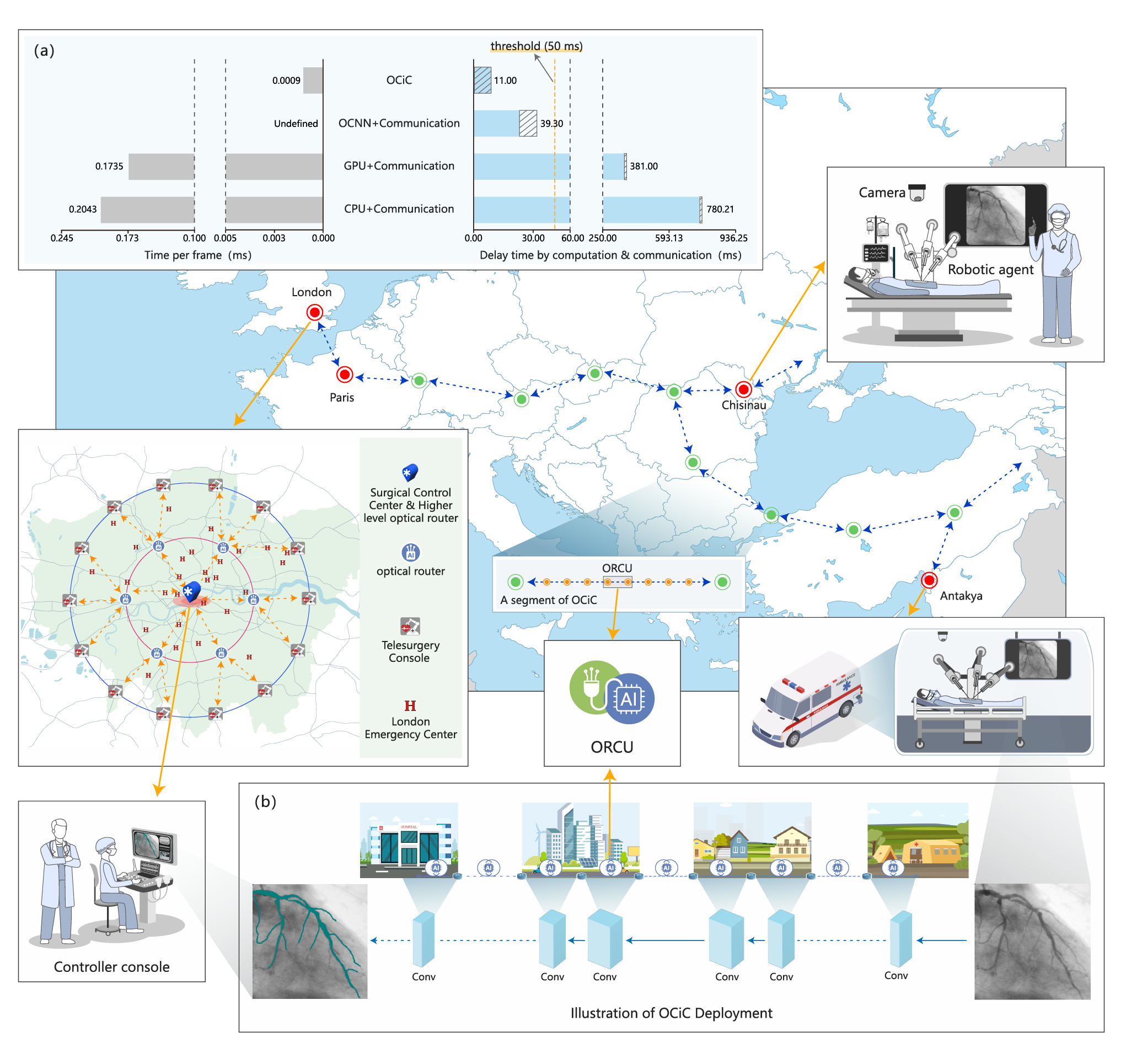}
    \caption{\textbf{The OCiC framework enables real-time, AI-enhanced perception for telesurgical procedures.}
    It addresses the lack of local sensory feedback that hampers intraoperative decision-making, while reducing cumulative latency from communication and inference. Panel (b) shows the OCiC architecture, consisting of geographically distributed, cascaded ORCUs that hierarchically implement distinct convolutional layers via fibre-based photonic computation. This design allows OCiC to perform AI inference concurrently with optical data transmission, effectively overlapping computation time with communication delays. As a result, the system supports AI-assisted telesurgery across intercontinental distances, remaining well within the clinically accepted one-way latency threshold of 50 ms. Panel (a) compares this approach to conventional systems, which rely on local computation followed by data transmission, showing that OCiC achieves higher frame rates and lower end-to-end latency (see Supplementary Information C.1 for further details). 
    Beyond latency reduction, OCiC enables dynamic redistribution of surgical capacity from well-resourced centres to underserved regions throughout its deployment network. This helps mitigate surgical delays due to limited specialist availability and restricted access to medical transport, especially in low- and middle-income countries, rural areas, and sparsely populated zones.
    The framework’s scalability is especially advantageous in disaster relief scenarios, where temporary OCiC branches can be rapidly deployed from nearby nodes, enabling immediate remote surgical support from clinical centres worldwide. }
    \label{fig: figure 1}
\end{figure*}

\begin{spacing}{0.9}

Telesurgery offers a compelling strategy to reduce global disparities in surgical care, addressing an unmet need affecting billions of people lacking timely and life-saving interventions \cite{aashna2022, shrime2015}. This challenge is most acute in underserved and remote regions, where establishing and sustaining a qualified surgical workforce remains prohibitively slow and costly \cite{verguet2015}. Low- and middle-income countries (LMICs) exemplify this crisis, experiencing severe shortages of surgical specialists, with densities as low as 0.7 per 100,000 people, compared to 56.9 per 100,000 in high-income regions \cite{holmer2015}. As a result, up to 94\% of individuals in these areas lack access to safe surgical care \cite{marlin2015}. Remote surgery thus represents a scalable solution to this global healthcare gap \cite{florian2025}, as initially demonstrated by pioneering achievements in telesurgery \cite{marescaux2001transatlantic, anvari2005}. 
However, wider adoption is limited by the lack of immediate, on-site sensory feedback  \cite{florian2025}, which complicates intraoperative decision-making and can compromise safety in high-precision procedures.

\ac{AI} is emerging as a transformative tool, demonstrating significant potential to improve intraoperative perception, decision-making, surgical precision, and operative workflows \cite{knudsen2024clinical}. These capabilities span diverse applications, including reconstructing 3D tissue structures from laparoscopic views \cite{huang2022self}, real-time tracking and localisation of surgical instruments \cite{zheng2019real}, and intraoperative planning assistance for vascular procedures \cite{zheng2019towards}. Despite these advances, current \ac{AI}-assisted robotic surgical systems predominantly rely on local electronic inference, introducing computational latency that frequently reaches hundreds of milliseconds \cite{de2023improving}. When combined with inherent communication delays in telesurgery, total latency often surpasses the clinically acceptable 100 ms round-trip threshold \cite{florian2025}, compromising the real-time control accuracy essential for intraoperative safety. This challenge is especially acute in latency-sensitive, high-precision procedures such as percutaneous coronary interventions \cite{patel2019long}, retinal microsurgery \cite{edwards2018first}, and neurosurgery \cite{jin2020comparative}. Consequently, compounded latency significantly impairs real-time surgical judgment and control \cite{yip2023artificial}, jeopardising patient safety and limiting the widespread adoption of robotic surgery in remote or resource-limited settings.

Here, we introduce an \ac{OCiC} framework that addresses latency constraints by fundamentally fusing deep-learning inference within the optical communication pathway. The system comprises a cascade of geographically distributed \acp{ORCU}, each hierarchically implementing a distinct convolutional layer through \ac{2D} photonic computation, as illustrated in Fig. \ref{fig: figure 1}b. This architecture intrinsically merges inference latency with optical transmission, significantly minimising total end-to-end delay, as shown in Fig. \ref{fig: figure 1}a. Beyond latency reduction, each \ac{ORCU} achieves high spectral efficiency and computational fidelity through ultra-stable kernels that maintain GPU-level accuracy under moderate optical noise. This design inherently mitigates cumulative error propagation \cite{Fu2024}, a longstanding challenge for scaling deep optical networks. Crucially, this electronic-level fidelity is achieved exclusively through architectural design, eliminating the need for auxiliary hardware \cite{Hamerly2022} or additional model retraining \cite{Xue2024,Wang2025,Bandyopadhyay2024}. Unlike conventional photonic processors constrained to centralized, on-site deployment \cite{feldmann2021parallel, xu202111, meng2023compact, dong2024partial, ahmed2025universal, ashtiani2022chip, dan2025}, the \ac{OCiC} framework is inherently compatible with existing optical network infrastructure, providing a scalable, in-network photonic \ac{AI} solution optimally tailored to latency-sensitive applications.

We demonstrate experimentally that a single ORCU can process up to 192 programmable kernels in parallel on one channel, which serves as the computational foundation for deep in-network optical inference. This configuration achieves a peak throughput of 69 \ac{TOPS} per channel, while maintaining inference fidelity within 0.1\% of the CPU baseline on the MNIST benchmark. To evaluate scalability and clinical applicability, we implemented a simplified \ac{OCiC} setup comprising cascaded \acp{ORCU} for coronary angiography segmentation, replacing several convolutional layers of the network. 
The system maintained inference fidelity within 0.1\% of GPU-based results under both controlled laboratory conditions and across a 38.9 km fibre link of the UK’s National Dark Fibre Facility (NDFF), despite variable real-world environmental conditions. 
These results demonstrate the \ac{OCiC} framework’s potential for robust, real-time medical \ac{AI} inference, paving the way for globally scalable telesurgery that meets clinical standards of precision and responsiveness.


\begin{figure*}[ht!]
    \centering
    \includegraphics[width=1.0\linewidth]{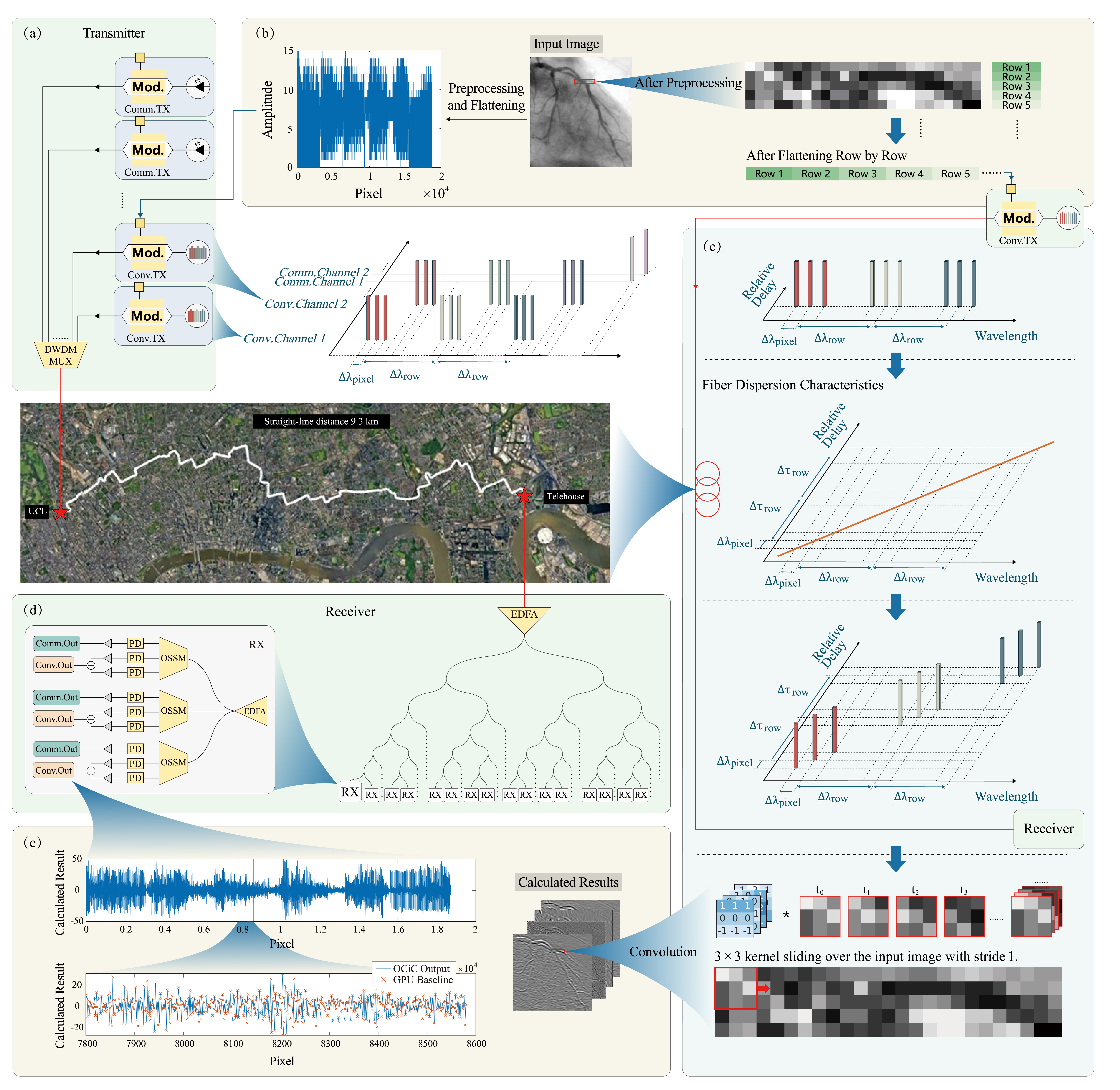}
    \caption{\textbf{\ac{ORCU} architecture enabling remote optical computation within communication fibres. } 
    (a) \ac{SMF} supports wavelength-division multiplexing, enabling simultaneous optical communication and computation. Communication channels transmit original images and residual data, while dedicated computational channels use two-dimensional structured optical combs to perform photonic convolutions over geographical distances. A representative 3×3 convolutional comb comprises three groups of three spectral modes, with wavelength spacings corresponding to pixel and row intervals. 
    (b) Two-dimensional input images are preprocessed and flattened into row-wise one-dimensional sequences before optical modulation and transmission. 
    (c) Leveraging linear fibre dispersion, computational channel wavelengths are temporally interleaved with precise delays, directly performing two-dimensional convolutions during optical transmission. 
    (d) At the receiver, a fan-out coupler array splits signals into 192 parallel optical paths. \acp{OSSM} assign kernel weights via spectral shaping; photodiodes convert and simultaneously sum optical signals into electrical outputs. 
    (e) Experimental convolution results show a high degree of consistency with theoretical predictions
    , illustrated alongside corresponding computed feature maps. Detailed methodological descriptions are provided in ~\hyperref[sec:method]{Methods}. More experimental results are provided in Supplementary Information B.}
    
    \label{fig: ORCU architecture}
\end{figure*}

\section*{\ac{ORCU} Enables Scalable Remote Photonic Computation}

To achieve practical scalability of the \ac{OCiC} framework, we developed the \ac{ORCU} architecture to address three critical deployment challenges: inference fidelity, spectral efficiency, and environmental robustness. As illustrated in Fig.\ref{fig: ORCU architecture}, \ac{ORCU} integrates optical communication and computation within a single fibre via wavelength-division multiplexing. Each computational channel executes massively parallel 2D photonic convolutions directly during optical transmission. Experimental validation confirms \ac{ORCU} achieves CPU/GPU-level inference fidelity, delivering a peak throughput of 69 \ac{TOPS} per channel and a spectral efficiency of 197.1 \ac{TOPS}/THz. Moreover, robust performance demonstrated in outdoor fibre deployments highlights \ac{ORCU}’s suitability as a scalable, geographically distributed optical platform. These findings position \ac{ORCU} as a promising solution for real-time, in-network photonic AI, particularly optimised for latency-sensitive applications.


Optical processors typically perform well on shallow benchmarks such as MNIST, yet commonly exhibit accuracy gaps of approximately 1–2\% relative to their electronic counterparts \cite{xu202111,meng2023compact,dong2024partial,ashtiani2022chip,dan2025}. This discrepancy intensifies in deeper networks primarily due to cumulative kernel noise, progressively degrading inference quality and potentially causing model collapse \cite{Fu2024}. Unlike transient feature noise suppressed by residual connections and nonlinear activations, kernel noise persists and accumulates throughout the network. To address this, \ac{ORCU} employs an ultra-stable 2D optical comb source specifically engineered to preserve kernel fidelity. Experimental validation confirms comb-line power variations consistently remain below 0.1 dB under laboratory conditions and below 0.2 dB during outdoor deployment (Supplementary Video). This stability surpasses that of widely used microring-based comb sources \cite{feldmann2021parallel,xu202111}, enabling ORCU to maintain CPU/GPU-level inference fidelity even under moderate optical noise conditions.


Another key advantage of ORCU is its exceptionally high computational spectral efficiency, critical for remote computing. Conventional fibre-based photonic accelerators rely on 1D convolutions to approximate inherently 2D operations. In contrast, ORCU directly implements parallel 2D convolutions (Fig.\ref{fig: ORCU architecture}c). Using a coupler array and receiver-side post-modulation (Fig.\ref{fig: ORCU architecture}d), a single 350 GHz channel can execute up to 192 programmable 3×3 kernels simultaneously, with a peak throughput of 69 TOPS. As one of the pioneering demonstrations of fibre-type photonic convolution accelerators \cite{xu202111}, the benchmark system requires 3.6 THz of optical bandwidth to reach 11.3 TOPS in 1D convolutions, equivalent to 3.8 TOPS for 3×3 kernels. By contrast, ORCU attains a spectral efficiency of 197.1 TOPS/THz, over 186 times higher than the 1.06 TOPS/THz reported in \cite{xu202111}. Compared with the representative optical cloud-computing approach described in \cite{Xing2025Seamless}, ORCU markedly improves spectral efficiency while avoiding the latency, overhead and synchronisation penalties of intermediate signal demodulation.

In addition to inference, \ac{ORCU} supports a computation-as-communication paradigm, replacing dedicated optical interconnections by directly embedding computation into the optical data path. This approach significantly reduces intermediate data transmission between nodes during collaborative inference, thereby improving spectral efficiency in geographically distributed learning systems. The resulting communication spectral efficiencies reach 87.8 bit/s/Hz for 4-bit resolution inputs and 153.6 bit/s/Hz for 7-bit, surpassing existing optical transmission records of 10.7 bit/s/Hz for single-mode
fibre (SMF) \cite{puttnam2024} and 51 bit/s/Hz for multi-core fibre \cite{puttnam2022}. 
These advancements position \ac{OCiC} as particularly valuable for privacy-sensitive, regulation-compliant multilayer split learning, especially in healthcare, finance, and the public sector.


To ensure robust performance under practical conditions, \ac{ORCU} integrates mechanisms specifically designed to mitigate environmental variability inherent in outdoor fibre deployments. A receiver-side post-modulation kernel assignment scheme enables real-time monitoring and dynamic compensation of power distortions induced by system nonlinearities and external perturbations. 
Furthermore, remote optical computing schemes may be affected by synchronisation issues arising from fibre-length fluctuations
\cite{hamerly2024netcast,sludds2022delocalized}. In contrast, \ac{ORCU} avoids these problems by performing 2D convolution entirely within the fibre through wavelength-to-time mapping. Collectively, these innovations substantially enhance system stability and reliability, underscoring \ac{ORCU}’s readiness for real-world deployment.


In summary, the \ac{ORCU} architecture achieves high-throughput optical inference at CPU/GPU-level fidelity, exceptional spectral efficiency, and robust performance across diverse environmental conditions within standard fibre infrastructure. Taken together, its inherent remote-computing capability positions \ac{OCiC} as a globally scalable platform for real-time, low-latency AI, particularly suitable for telesurgical applications.

\begin{figure*}[!ht]
    \centering
    \includegraphics[width=0.950\linewidth]{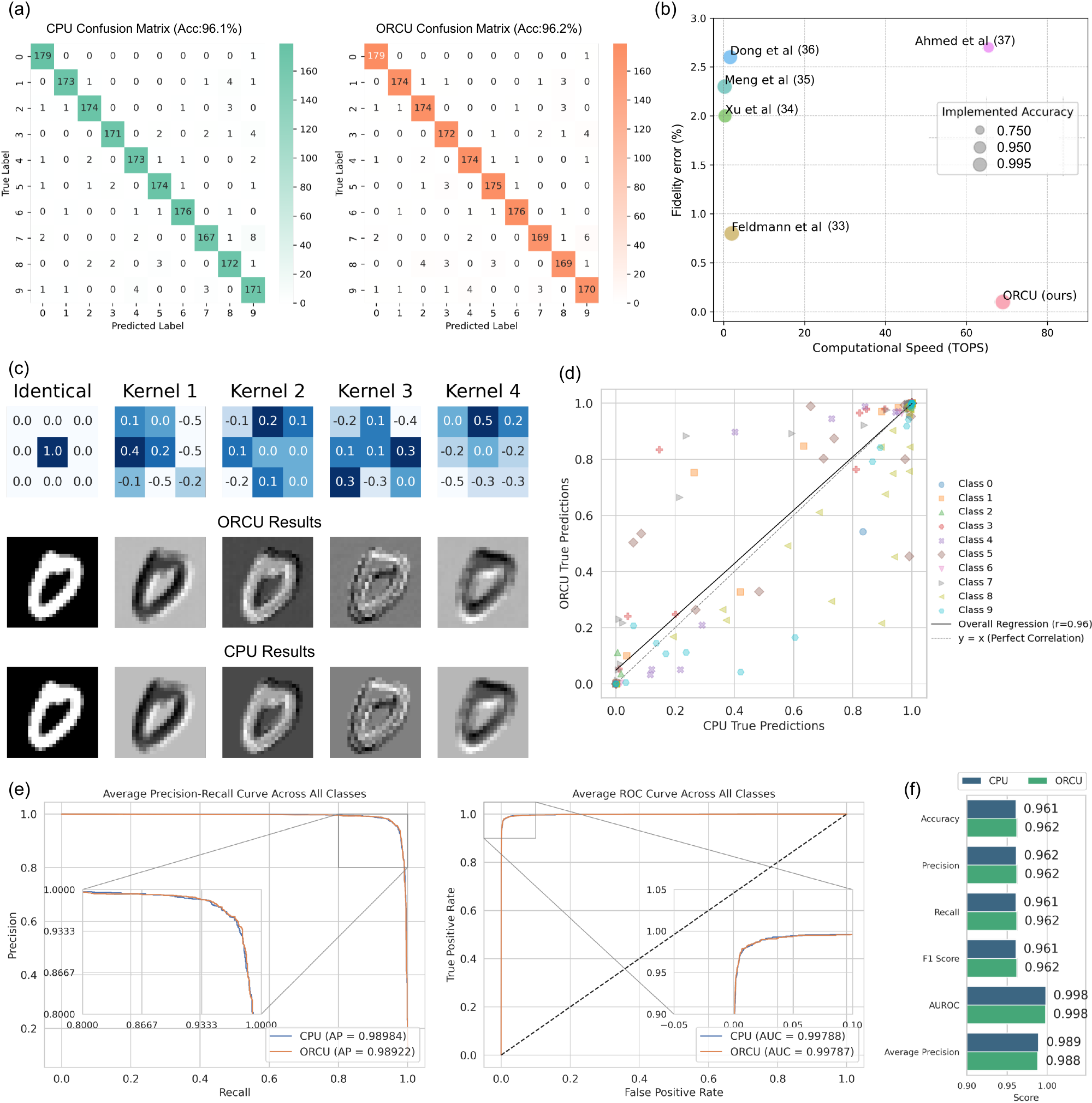}
    \caption{
    \textbf{\ac{ORCU} achieves CPU-level inference accuracy with superior fidelity on the MNIST classification task.}
    (a) Confusion matrices demonstrate nearly identical classification accuracy between \ac{ORCU} (96.2\%) and CPU-based inference (96.1\%).
    (b) Bubble chart comparing fidelity error (accuracy deviation from electronic inference) and computational speed highlights \ac{ORCU}’s enhanced fidelity relative to benchmark optical processors \cite{feldmann2021parallel,xu202111,meng2023compact,dong2024partial,ahmed2025universal}.
    (c) Convolutional feature maps generated by four pre-trained kernels confirm that \ac{ORCU} outputs closely match CPU-computed results at the feature-extraction level.
    (d) Scatter plot of prediction probabilities from \ac{ORCU} versus CPU computations exhibits high consistency across all classes (regression correlation, r = 0.96).
    (e) Precision–Recall (left) and receiver operating characteristic (\ac{ROC}, right) curves indicate robust classification performance across a wide range of decision thresholds.
    (f) Bar charts summarising key performance metrics—accuracy, precision, recall, F1 score, \ac{AUROC}, and average precision—demonstrate that \ac{ORCU} matches or surpasses CPU-based performance.
    Together, these results confirm \ac{ORCU}’s capability to achieve CPU-level inference accuracy while supporting high-throughput parallel optical processing, positioning it as a promising candidate for scalable optical computing architectures.
    }
    \label{fig: mnist_results}
\end{figure*}

\section*{\ac{ORCU} Delivers High-Throughput Optical Inference with CPU-level Fidelity}
We benchmarked \ac{ORCU}’s inference fidelity against a CPU baseline using the MNIST classification task. Four pre-trained 3×3 convolution kernels from a \ac{CNN} were mapped onto optical filters, each randomly assigned to one of \ac{ORCU}’s 192 programmable parallel-processing kernels. 

The confusion matrices (Fig.\ref{fig: mnist_results}a) and feature maps (Fig.\ref{fig: mnist_results}c) produced by \ac{ORCU} closely matched CPU results, demonstrating no discernible qualitative differences. \ac{ORCU} achieved a classification accuracy of 96.2\% on an identical 1,800-image test set, closely matching the CPU benchmark (96.1\%). This minimal inference discrepancy is further supported by strong agreement between experimental and theoretical waveforms and their prediction outcomes (see Supplementary Information B.1-3). Additional fidelity indicators include a regression correlation of 0.96 (Fig.\ref{fig: mnist_results}d), precision–recall and receiver operating characteristic (ROC) curves (Fig.\ref{fig: mnist_results}e), and metrics such as F1 score, area under the ROC curve (AUROC), and average precision (Fig.\ref{fig: mnist_results}f). 

Collectively, these results confirm \ac{ORCU} achieves CPU-level inference accuracy while enabling high-throughput parallel processing, surpassing existing benchmark photonic processors (Fig.\ref{fig: mnist_results}b). This high fidelity allows \ac{ORCU} to effectively mitigate cumulative errors even under moderate optical noise, providing a robust foundation for scalable, deep-layered \ac{OCiC} deployments across geographically diverse regions.

\begin{figure*}[!ht]
    \centering
    \includegraphics[width=1.0\linewidth]{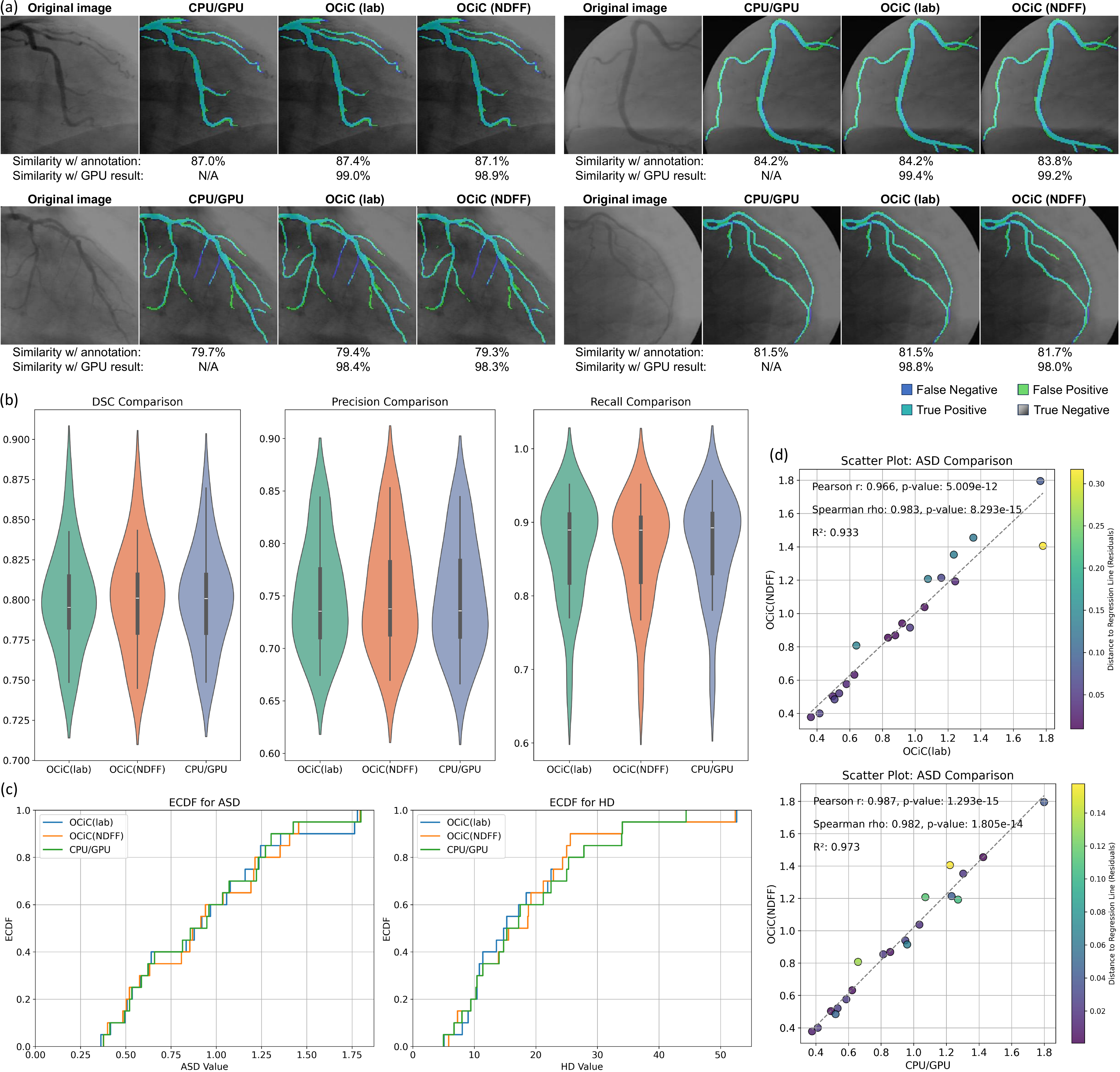}
    \caption{
    \textbf{\ac{OCiC} achieves high-fidelity, real-time vascular segmentation across laboratory and field deployments. } 
    (a) Representative X-ray angiograms and corresponding coronary artery segmentation results from the GPU baseline, \ac{OCiC} (lab), and \ac{OCiC} (\ac{NDFF}). The segmentation outputs from both \ac{OCiC} implementations closely match the GPU reference, confirming high inference fidelity. 
    (b) Violin plots summarising the \ac{DSC}, precision, and recall across 20 unseen angiograms. \ac{OCiC} (lab) and \ac{OCiC} (\ac{NDFF}) exhibit performance statistically indistinguishable from the GPU baseline. 
    (c) \acp{ECDF} of \ac{ASD} and \ac{HD}, showing close spatial agreement between \ac{OCiC}-derived segmentations and the GPU reference. 
    (d) Scatter plots of \ac{ASD} measurements confirm strong correlations among the GPU, \ac{OCiC} (lab), and \ac{OCiC} (\ac{NDFF}). Outliers are indicated in yellow.
    }
    \label{fig: seg_results}
\end{figure*}

\section*{\ac{OCiC} Enables Environmentally Robust, AI-enhanced Perception for Telesurgeries}

A major challenge in telesurgery is the lack of immediate, high-fidelity sensory feedback due to network latency and compression, which complicates procedures requiring precise, real-time decisions.
This limitation remains a critical barrier to the broader adoption of telesurgery, particularly in anatomically complex scenarios where even minor errors or delays can have life-threatening consequences. Nonetheless, remote interventions for acute vascular emergencies such as myocardial infarctions and strokes remain urgently needed due to their narrow therapeutic windows \cite{bergmark2022acute, powers2020acute} and a global shortage of interventional cardiologists and cardiac specialists \cite{marlin2015, vervoort2020global, narang2016supply}. 
By leveraging \ac{ORCU}’s CPU/GPU-level inference fidelity and its intrinsic remote-computing capability, the \ac{OCiC} framework delivers real-time, \ac{AI}-enhanced perception without introducing additional computational latency. This approach will simplify remote surgical interventions and significantly expand timely access to specialised care, especially in underserved regions.


To evaluate \ac{OCiC}’s clinical inference capabilities under real-world deployment conditions, we implemented key convolutional layers from a pre-trained 161-layer U-DenseNet for coronary artery segmentation in contrast-enhanced X-ray angiograms. During vascular interventions, real-time fluoroscopic imaging provides essential intraoperative guidance for precise localisation and treatment of vascular pathologies. 
By delivering real-time, AI-enhanced, fluoroscopy-synchronised visualisation, OCiC strengthens visual feedback in the absence of immediate haptic cues, thereby supporting more precise and timely decision-making.
While demonstrated here for vascular surgery, this AI-assisted approach has broad applicability and can be seamlessly extended to other surgical domains, such as endoscopic interventions, further enhancing the quality and reliability of remote procedures. 
To assess system robustness, the \ac{ORCU} was specifically assigned to compute the first and last convolutional layers. These layers are particularly sensitive to distinct noise types: kernel noise in the initial layer propagates through subsequent layers, while feature noise in the final layer directly impacts segmentation accuracy. Performance was evaluated on 20 unseen angiographic images across three scenarios: (1) GPU-only baseline, (2) \ac{OCiC} in a controlled laboratory environment, and (3) \ac{OCiC} deployed over a 38.9 km outdoor dark fibre link (\ac{NDFF}) to assess robustness under real-world conditions. Comprehensive details of training methods and deployment configurations are provided in ~\hyperref[sec:method]{Methods}. 


As illustrated in Fig.~\ref{fig: seg_results}a, the \ac{OCiC} framework demonstrated robust and consistent segmentation performance under both controlled laboratory and outdoor \ac{NDFF} conditions. The system achieved nearly identical average Dice scores of 0.800 (laboratory) and 0.801 (\ac{NDFF}), closely matching the GPU baseline of 0.801. Using GPU-derived predictions as a reference standard, F1 score analysis yielded values of 0.984 (laboratory) and 0.983 (\ac{NDFF}), indicating minimal deviation from GPU-based inference. Violin plots of Dice scores, precision, and recall (Fig.~\ref{fig: seg_results}b) confirmed that distributions were statistically indistinguishable from GPU-based results, highlighting the system’s stability. Empirical Cumulative Distribution Function (ECDF) (Fig.~\ref{fig: seg_results}c) and scatter plots (Fig.~\ref{fig: seg_results}d) further confirmed minimal output discrepancies, while strong agreement in Average Surface Distance (ASD) and Hausdorff Distance (HD) metrics underscored the preservation of spatial segmentation accuracy. Comprehensive results for all 20 test cases, including vascular segmentation maps, per-image Dice scores, and pixel-level comparisons, are provided in the Supplementary Information B.4-6. 
Collectively, these results show that OCiC achieves real-time inference with performance comparable to a GPU baseline.
Its reliable accuracy under both laboratory and realistic field conditions further demonstrates resilience to environmental perturbations, supporting its suitability for latency-sensitive applications such as telesurgery.

\section*{Discussion and Conclusion}

The \ac{OCiC} framework leverages the \ac{ORCU}’s capabilities to reduce error accumulation and perform \ac{AI} inference concurrently with optical data transmission. This approach enables the direct deployment of pre-trained models, specifically tailored to surgical tasks, over existing fibre-optic infrastructure. The architecture provides real-time, high-fidelity perception for surgeons during complex telesurgical procedures, substantially reducing latency-related risks and potentially enhancing patient safety.

First, experimental results demonstrate strong quantitative agreement with electronic inference, underscoring \ac{ORCU}’s robustness against error accumulation, which is a crucial prerequisite for deploying deep learning networks on optical hardware \cite{Fu2024}. To rigorously quantify this robustness, we conducted noise-injection simulations targeting two dominant sources: feature noise, arising from transient optical and electrical fluctuations, and kernel noise, induced by system instabilities (see Extended Data Fig.~\ref{fig:figure5}). Using a conservative accuracy margin of 0.1\% derived from the MNIST benchmark, we estimated the potential accuracy degradation resulting from full-network deployment of the surgical segmentation model within \ac{OCiC}. Under these conditions, the projected full-network accuracy ranged from 0.80 (feature noise) to 0.79 (kernel noise), preserving 98.7–99.8\% of the baseline accuracy (0.801). These results strongly support the feasibility of directly deploying pre-trained deep neural networks within \ac{OCiC}, delivering real-time, AI-enhanced perception to assist intraoperative decision-making during telesurgical procedures. 
Crucially, our simulations indicate that kernel instability poses a significantly greater threat to inference accuracy than feature noise. Even modest degradation observed in relatively simple tasks (e.g., 1.1\% on MNIST) can escalate beyond 10\% in deeper, more complex networks, severely constraining the scalability of photonic AI systems. This discrepancy arises from a fundamental distinction: kernel noise disrupts learned weights, causing structural perturbations that propagate across layers and progressively degrade feature extraction. In contrast, feature noise is transient and typically attenuates as it propagates, provided it remains moderate and unbiased. Collectively, these findings underscore the critical need for improving stability and precision in photonic components, particularly microring resonators and phase-change materials, thereby charting a clear roadmap towards robust deep-layer inference in future optical AI systems.

Second, the \ac{OCiC} framework strategically overlaps computational processing with optical transmission, substantially minimising end-to-end latency and extending the geographical reach of real-time, AI-assisted telesurgical interventions. Given a typical propagation speed of approximately $2\times10^8$ m/s in optical fibre, the clinically acceptable round-trip latency threshold of 100 ms \cite{florian2025} translates into a theoretical operational range of approximately 10,000 km. This operational range comfortably covers key global routes such as London to Tokyo (9,600 km), Los Angeles to Buenos Aires (9,800 km), and Paris to Cape Town (9,300 km), underscoring the OCiC system’s architectural suitability for latency-critical, high-precision interventions across continents. Crucially, the framework consistently maintains latency within clinically acceptable limits, facilitating seamless, real-time collaboration among clinical centres worldwide and integrating distributed global expertise \cite{Rogers2024}. 
Experimentally, a single computational channel demonstrated a peak throughput of 69 \ac{TOPS}. Leveraging the full 11.9 THz optical bandwidth across the C and L bands, each \ac{ORCU} can concurrently process up to 34 parallel computational channels per \ac{SMF}, significantly enhancing batch-processing capacity. In principle, a single 10-km bidirectional \ac{SMF} segment can deliver a peak throughput of up to 4.7 peta-operations per second. Extended to a 10,000-km \ac{OCiC} network, the theoretical peak throughput scales markedly, potentially reaching 4.9 exa-operations per second using only one fibre per \ac{ORCU} segment. In practical deployments, standard optical cables containing hundreds of fibres enable orders-of-magnitude greater distributed photonic computing capacity. Such infrastructure effectively transforms conventional long-haul optical links into continental-scale computing fabrics, offering shared, decentralised AI resources accessible to urban, suburban, and rural communities alike. These capabilities lay the foundation for global-scale, AI-assisted telesurgical interventions and facilitate real-time, cross-border experimental collaboration among geographically dispersed clinical, research, and academic institutions. Potential applications include multi-centre clinical trials, remote participation in international physics experiments, and seamless cross-border integration of scientific data and knowledge.

Deploying remote operating rooms along the \ac{OCiC} network in LMICs, rural communities, and emergency-affected regions can deliver timely, life-saving surgical interventions to underserved populations. This model directly addresses urgent healthcare needs in resource-limited settings without requiring long-term investments in local specialist training, while facilitating effective redistribution of surplus surgical capacity from well-resourced centres to regions with limited access. Ultimately, it presents a cost-effective, scalable, and equitable strategy to reduce global health disparities and accelerate progress towards achieving the Global Surgery Goals \cite{anna2014}.

\end{spacing}

\begin{spacing}{0.05} 
    \bibliography{ref}   
  \end{spacing} 
\clearpage

\section*{Methods}
\label{sec:method}

\setcounter{figure}{0}
\renewcommand{\figurename}{Extended Data Fig.}

\subsection*{Principle of Proposed \ac{ORCU}}

\acp{CNN} are a class of deep learning architectures widely applied to visual data processing. At their core, convolutional layers employ small learnable kernels to extract local features such as edges and textures from input images. As each kernel convolves across the image, the resulting feature map encodes distinct spatial patterns through its aggregated responses. Multiple kernels operate in parallel, enabling the network to capture diverse visual cues that underpin its success in image recognition and related tasks.

\begin{figure*}[!ht]
    \centering
    \includegraphics[width=0.9\linewidth]{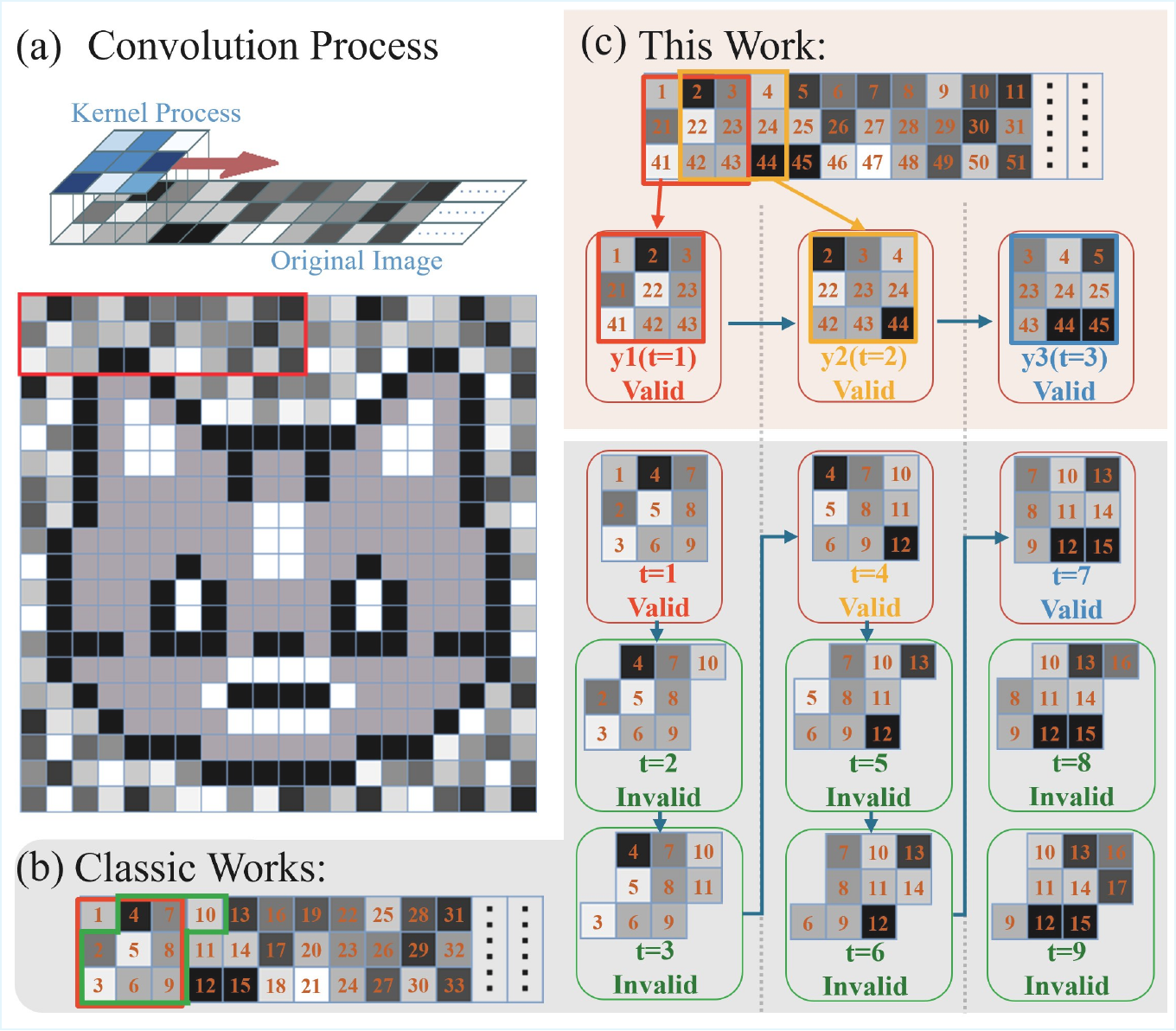}
    \caption{\textbf{Illustration of our 2D photonic convolution.}
    (a) In 2D convolution, a kernel (e.g., 3 × 3) is placed at the top-left corner of the input, where it multiplies element-wise with the overlapping region, sums the results, and adds a bias to produce a single output. The kernel then slides across the input from left to right, row by row, until all valid regions are covered. Each result is assigned to the corresponding position in the output feature map.
    (b) In a conventional optical convolution processor, pixels are fed in column-major order: for each column, typically only the first three pixels are input sequentially from top to bottom before moving to the next column. Such methods enable a one-dimensional photonic convolution processor to handle 2D images, but introduce an inherent stride limitation, reducing the effective throughput to one-third of the nominal rate for 3 × 3 convolutions. As illustrated in the right panel of (b), only outputs at $t_{1}$, $t_{4}$, $t_{7}$, … are valid, while those at $t_{2}$, $t_{3}$, $t_{5}$, $t_{6}$, $t_{8}$, $t_{9}$, … are redundant. 
    (c) In contrast, our design supports 2D photonic convolution with row-by-row pixel feeding into the optical link, thereby achieving nearly 100\% computational efficiency.
    }
\label{fig: 2D_convolution}
\end{figure*}

The computational demands of large-scale CNNs are driving the development of photonic tensor accelerators that exploit the high bandwidth of optical systems for parallel convolution operations. In fibre-based designs, two-dimensional images are typically flattened into one-dimensional data streams. These are then transmitted into optical systems via a \ac{MZM}. However, a stride limitation inherent to most fibre-type photonic accelerators substantially reduces the effective computing speed, achieving only one-third of the nominal rate for 3×3 convolutions
, as illustrated in Extended Data Fig.~\ref{fig: 2D_convolution}. To overcome this, \ac{ORCU} employs comb sets with an engineered frequency-domain distribution as the light source for each computation channel, enabling direct \ac{2D} convolution. For a 3×3 kernel, the system generates an unfolded comb set of three sparsely spaced groups, each corresponding to one row of the kernel. Within each group, three densely spaced comb lines represent the elements of that row, separated by $\lambda_{\text{pixel}}$. This wavelength spacing corresponds to the one-pixel time delay introduced via wavelength–to–time mapping in the \ac{SMF}. The comb groups are separated by a larger wavelength interval, $\lambda_{\text{row}}$, providing temporal separation between adjacent pixel rows. Fig.\ref{fig: ORCU architecture}a shows multiple computation and communication channels sharing the same \ac{SMF}, with a single ultra-broad comb (typically generated by a mode-locked laser or microring resonator) divided among all channels. Experimental measurements indicate that comb stability, particularly in power fluctuation and wavelength spacing, exerts a stronger influence on performance than optical noise. Nevertheless, generating ultra-stable, high-repetition-rate combs with microring resonators or mode-locked lasers remains technically challenging. In this work, we adopt electro-optic comb generation with modulated coupled laser sources to ensure the stability required for high-performance photonic computing.

The preprocessing stage includes flattening \ac{2D} images into one-dimensional row-wise sequences (Fig.\ref{fig: ORCU architecture}b) and applying predistortion (see Supplementary Information C.2 for details). After preprocessing, the optical signal in each computational channel is modulated by the flattened image at a symbol rate $R_s$ (symbols per second) and transmitted through a \ac{SMF} of approximately 10 km. As shown in Fig.\ref{fig: ORCU architecture}c, the \ac{SMF} performs wavelength–to–time mapping, interleaving the different wavelengths of each computational channel with the required time delays. In this configuration, adjacent comb lines correspond to neighbouring pixels, while the spacing between different comb groups represents the time delay between image rows. At the \ac{SMF} output, the wavelength–time interleaved data streams yield unweighted, wavelength-domain kernels. These kernels directly perform \ac{2D} convolution, processing data on a pixel-by-pixel basis over successive symbol durations (lower panel of Fig.\ref{fig: ORCU architecture}c). This direct \ac{2D} convolution also mitigates the stripe problem common in other fibre-based photonic accelerators, a previously described limitation that significantly reduces effective computing speed (Extended Data Fig.~\ref{fig: 2D_convolution}).

Additionally, the proposed \ac{ORCU} uses a post-modulation strategy for kernel assignment (Fig.\ref{fig: ORCU architecture}d) to enhance its capability for parallel convolutional computation. At the receiver end of the \ac{SMF}, the unweighted \ac{2D} kernels are replicated into multiple copies using a coupler array, with each copy individually shaped by an \ac{OSSM}. This configuration improves spectral efficiency and increases the practical value of the technique for deep network deployment. Experimental results show that the current \ac{ORCU} design achieves high computational efficiency while maintaining CPU-level inference accuracy. It supports the parallel computation of up to 192 kernels ($2^6$ × 3 = 192 kernel paths) on a single computational channel using only 350.7 GHz of bandwidth. This corresponds to a total computing power of 69.12 \ac{TOPS} (192 × 2 × 9 × 20 GHz) and a computational spectral efficiency of 197.1 \ac{TOPS}/THz. The simultaneous arrival of convolutional results from multiple kernels at the receiver reduces the required refresh cycles to process the same image, thereby lowering storage pressure across the entire \ac{OCiC} link and minimising latency. Furthermore, the post-modulation approach enables real-time supervision and adjustment of kernel information, allowing the system to dynamically adapt to environmental fluctuations and transmitter-end variations.


\subsection*{Experiment Setups}


This study investigates whether the proposed \ac{ORCU} can mitigate structural perturbations in convolution kernels. Achieving this capability is essential for deploying optical computing units in deep learning networks compliant with \ac{AI} scaling laws. Simulations indicate that deep networks are substantially more sensitive to structural perturbations caused by kernel noise than to feature noise affecting the feature map. Stable comb sources and precise \acp{OSSM} are therefore critical to preserving kernel fidelity and system robustness. To ensure this stability, \ac{EO} comb generation using modulated coupled laser sources is employed. Experimental validation shows that \ac{EO} combs maintain power fluctuations below 0.1 dB in temperature-controlled environments and below 0.2 dB under room-temperature conditions with outdoor dark-fibre links, as shown in the Supplementary Video. In our proof-of-concept system, waveshapers are used as \acp{OSSM} to provide accurate and stable kernel modulation, thereby validating high-fidelity optical inference of \ac{ORCU} and enabling exploration of its practical performance limits. More integrated alternatives, such as microring arrays or arrayed waveguide gratings with variable optical attenuators, offer higher integration potential. However, their instability, arising from thermal drift and control complexity, undermines kernel fidelity and limits their practicality in deep learning deployment. Taken together, these results underscore that continued development of stable, integrated, and cost-effective comb sources and \acp{OSSM} is crucial for scalable optical computing systems to meet the escalating demands imposed by \ac{AI} scaling laws. 

We conducted three experiments, each with a tailored setup for specific objectives. The first experiment focused on performance verification. A large-scale fan-out coupler array was constructed using six layers of 1 × 2 couplers and one layer of 1 × 3 couplers. This array uniformly distributed the received unweighted convolutional signals into 192 optical sub-paths with comparable optical quality (power, signal-to-noise and distortion ratio, and distortion characteristics), thereby enabling parallel computation of 192 distinct convolution kernels under consistent conditions. To assess performance, four of the 192 sub-paths were weighted with convolution kernels from a pre-trained \ac{CNN} for the MNIST classification task. The optical system achieved classification accuracy (96.2\%) nearly identical to CPU-based computation (96.1\%). These results confirm that \ac{ORCU} sustains CPU-level inference fidelity under moderate optical noise, provided kernel stability and precision are maintained. 

In this MNIST classification experiment, a 9.5 km \ac{SMF} served as the core of \ac{ORCU}, providing stable, linear wavelength-to-time mapping \citemethods{yang2024photonic}. The flattened image sequence was transmitted at 20 GBaud. The \ac{EO} comb generator was realised by combining three \ac{CW} lasers (1560.757 nm, 1552.122 nm, and 1543.212 nm, with output powers of 15.5 dBm, 14.5 dBm, and 16.5 dBm) through a 3 × 1 coupler. The unequal power levels were intentionally set to compensate for the nonlinear response of the \acp{EDFA}. The combined optical signal then passed through an \ac{EO} comb generator consisting of two cascaded phase modulators driven by a 38.963 GHz RF sine wave, producing three groups of three-line combs with the required one-pixel wavelength spacing. The resulting nine comb lines were centred at 1542.903 nm, 1543.212 nm, 1543.522 nm, 1551.809 nm, 1552.122 nm, 1552.436 nm, 1560.441 nm, 1560.757 nm, and 1561.074 nm, mapping directly to the nine elements of a 3 × 3 convolution kernel. 

To achieve accurate wavelength–to–time interleaving, the one-pixel wavelength spacing, ($\lambda_\text{pixel}$), is defined as
\begin{equation}
\begin{aligned}
\lambda_{\text{pixel}} = \frac{\tau_{\text{pixel}}}{D \cdot L} = \frac{1}{D \cdot L \cdot R_S}
\end{aligned}
\end{equation}
where \textit{L} is the \ac{SMF} length, \textit{D} is the fibre dispersion, \textit{$R_{S}$} is the symbol rate of the input data, and the pixel duration is given by $\tau_{\text{pixel}}=\frac{1}{R_{S}}$.

\begin{figure*}[!ht]
    \centering
    \includegraphics[width=1.0\linewidth]{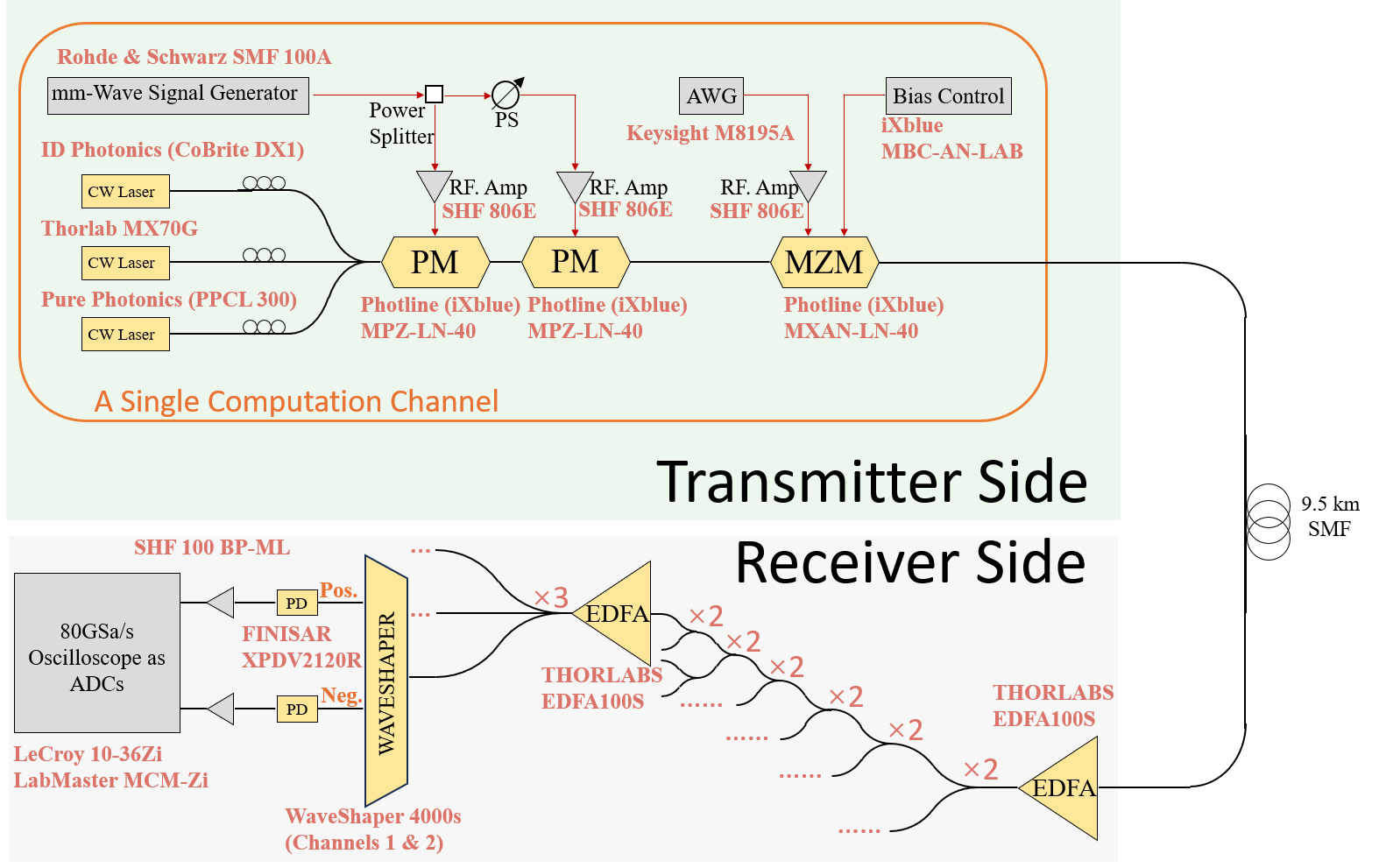}
    \caption{
Experimental arrangement for MNIST classification task
    }
    \label{fig: System_MNIST}
\end{figure*}

\vspace{0.5cm}
After propagation through the 9.5 km \ac{SMF}, the unweighted convolution signal is fed into the fan-out coupler array. Two \acp{EDFA} compensate for insertion and transmission losses in the transmitter, fibre, and fan-out stages. At the end of each optical sub-path, a waveshaper acting as the \ac{OSSM} modulates the power of each comb line to encode the corresponding convolution-kernel weights. Because optical power cannot directly represent negative values, positive and negative kernel components are modulated separately and delivered to the two output ports of the waveshaper. The final convolution result is obtained by subtracting the two outputs, which can be realised with a balanced photodiode, a differential amplifier, or, as implemented in this experiment, \ac{DSP}. The complete experimental setup, including device specifications, is illustrated in Extended Data Fig.~\ref{fig: System_MNIST}. 

The second experiment evaluates the capability of the proposed \ac{OCiC} to perform coronary artery segmentation in support of real-time \ac{AI}-assisted telestenting surgery. To assess whether the \ac{OCiC} system can execute this task using a pre-trained model, a 161-layer U-DenseNet previously trained on a GPU was employed. In this setup, \ac{ORCU} was used to execute the first and last convolutional layers. This choice reflects two considerations. The first layer is most susceptible to error accumulation due to kernel noise, which can propagate through the entire network. The last layer is highly sensitive to feature noise, directly impacting output accuracy. Given the structure of the U-DenseNet, the coupler array was scaled down to five layers, yielding 32 optical sub-paths, each corresponding to one of the 32 convolution kernels within each channel. Additionally, an extra communication channel was included in the experiment to assess the feasibility of residual data transmission through the \ac{OCiC} system. 

\begin{figure*}[!ht]
    \centering
    \includegraphics[width=1.0\linewidth]{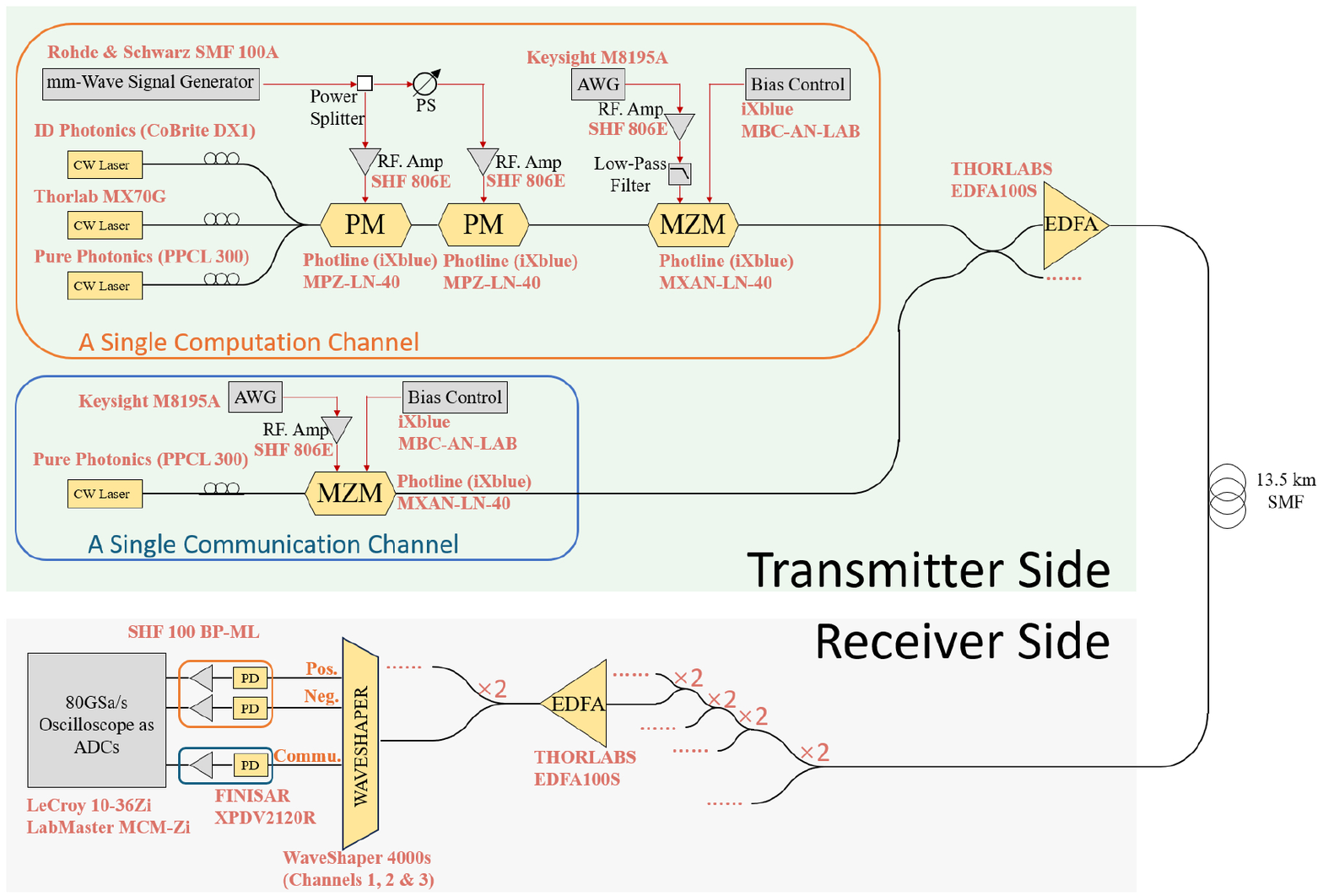}
    \caption{
Experimental arrangement for coronary angiography segmentation task in controlled laboratory settings
    }
    \label{fig: System_Segmentation}
\end{figure*}

To reflect the geographical context of London, the \ac{SMF} length for convolutional kernel construction and data transmission was extended to 13.5 km in this experiment. This configuration enabled both convolutional layers to be processed over a total distance of 27 km, sufficient to cover central London to its outskirts in a telesurgical scenario. To accommodate the longer fibre length, the data rate was reduced to 16 GBaud, and the comb generation frequency was adjusted to 34.72 GHz to provide the required one-pixel wavelength spacing. The testing dataset comprised 20 coronary artery X-ray images, each 128 × 128 pixels in size. Owing to the limited spectral bandwidth of the C+L band, each image was partitioned into six 24 × 128 sub-images, since the spectral spacing for 256 pixels (two rows) could not be supported simultaneously. Details of the image partitioning process are provided in Supplementary Information C.2. In this experiment, the 3 × 3 comb group was tuned to the following wavelengths: 1545.149 nm, 1545.426 nm, 1545.702 nm, 1551.844 nm, 1552.122 nm, 1552.402 nm, 1558.380 nm, 1558.661 nm, and 1558.943 nm, corresponding to the revised symbol rate, fibre length, and pixel count per row. After each convolutional operation, the sub-images were recombined to reconstruct the complete image. The full experimental setup is shown in Extended Data Fig.~\ref{fig: System_Segmentation}.   

\begin{figure*}[!ht]
    \centering
    \includegraphics[width=0.9\linewidth]{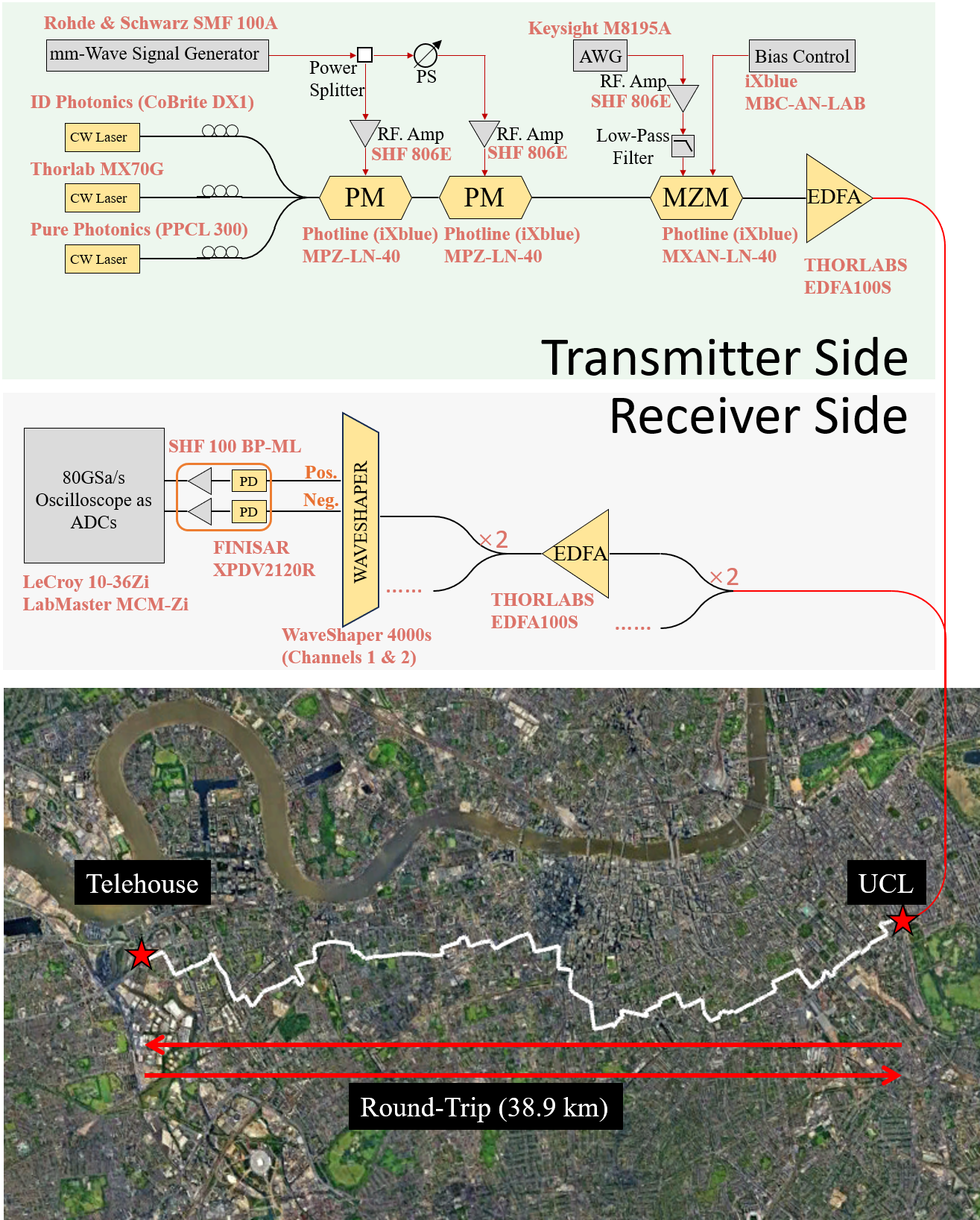}
    \caption{
Experimental arrangement for coronary angiography segmentation task with NDFF deployment round-trip length of 38.9 km. Map based on satellite imagery from Google Earth \protect\citemethods{GoogleEarth2025}.}
    \label{fig: System_NDFF}
\end{figure*}

Finally, the same coronary artery segmentation task was carried out over a 38.9 km segment of dark fibre within the UK’s \ac{NDFF} to assess the robustness of the proposed \ac{ORCU} under outdoor conditions. This experiment validated the \ac{OCiC} deployment under real-world conditions, encompassing temperature variations, weather changes, polarisation fluctuations, and physical disturbances from nearby traffic (temperature variations, weather changes are summarised Supplementary Information C.3). Owing to limited physical space at the \ac{NDFF} laboratory and an unexpectedly high fibre-optic power loss of 14 dB, a simplified configuration using a 1-to-4 coupler array was adopted, which was sufficient to perform the segmentation task. To accommodate the longer fibre, the symbol rate was reduced to 8 GBaud, and the 3 × 3 comb group was tuned to the following wavelengths: 1547.166 nm, 1547.364 nm, 1547.562 nm, 1551.923 nm, 1552.122 nm, 1552.322 nm, 1556.600 nm, 1556.801 nm, and 1557.001 nm. The full experimental setup is illustrated in Extended Data Fig.~\ref{fig: System_NDFF}.

Additionally, PAM-16 (16-level Pulse Amplitude Modulation) was selected as the modulation format for the MNIST classification task and the first layer of the coronary artery segmentation task, primarily to reflect practical real-world communication configurations. Widely adopted in experimental and research contexts, PAM-16 offers a balanced trade-off between modulation complexity and implementation feasibility. This choice avoids overly idealised assumptions, thereby enhancing the real-world relevance of both the system model and experimental results. In contrast, the final layer used unconstrained input levels. Both input data and kernel weights retained five significant digits of precision, rather than being quantised to 4-bit representation. This approach helps prevent artificial performance degradation and allows for meaningful comparison with GPU-based results. Moreover, since the input to the final layer includes negative values, it was split into two images: one containing only positive values, and the other the absolute values of negative components. These were processed sequentially using the same convolutional kernels, and the outputs were recombined via a \ac{DSP} programme.

\newpage

\newpage

\newpage

\subsection*{Network structure design and model training}
\label{sec:network_design}

We experimentally validated the \ac{OCiC} framework on two representative computer vision tasks: image classification and semantic segmentation. Image classification is based on MNIST hand-written digit dataset \citemethods{lecun1998mnist}, while semantic segmentation task is based on dataset from Cervantes et al. \citemethods{cervantes2019coronary}. Both pipelines followed a unified workflow of model construction, training, and inference, with selected inference stages executed on the \ac{OCiC} platform.

For the classification task, we constructed a \ac{CNN} with one convolutional layer (four kernels), a 16-neuron hidden layer, and a 10-neuron output layer corresponding to the MNIST digit classes. The model was trained on 20,000 MNIST images and tested on 1,800 held-out examples. During inference, the convolutional layer was replaced by \ac{ORCU}, where optically computed features were fed directly into the fully connected layers. The output was a 10-dimensional probability vector, with the highest-probability class selected as the prediction.

For semantic segmentation, representative of \ac{AI}-assisted telesurgical scenarios, we developed a deep neural network based on a hybrid DenseNet \citemethods{huang2017densely} and U-Net \citemethods{ronneberger2015u} architecture, referred to as U-DenseNet \citemethods{U-densenet}. Dense and transition blocks were arranged in a U-shaped topology to enable high-resolution segmentation of coronary vasculature. The model was trained on 107 contrast-enhanced X-ray angiograms and tested on 20 cases, with no test data used during training. During evaluation, the first and final convolutional layers were executed on the \ac{OCiC} system. Inference was validated under both controlled laboratory conditions and deployment across a 38.9 km segment of the UK’s \ac{NDFF}, with GPU-based inference as the baseline.

Both networks were trained on a high-performance computing server at University College London, equipped with 4 × Intel Xeon Gold 6126 CPUs (2.60 GHz), 4 × NVIDIA Tesla V100 GPUs (32 GB each), and 512 GB system RAM. Further architectural details are provided in the Supplementary Information A.4.

\subsection*{Noise Analysis}
\begin{figure}[h]
    \centering
    \includegraphics[width=0.9\linewidth, trim=1.2cm 0.8cm 0.6cm 0.4cm]{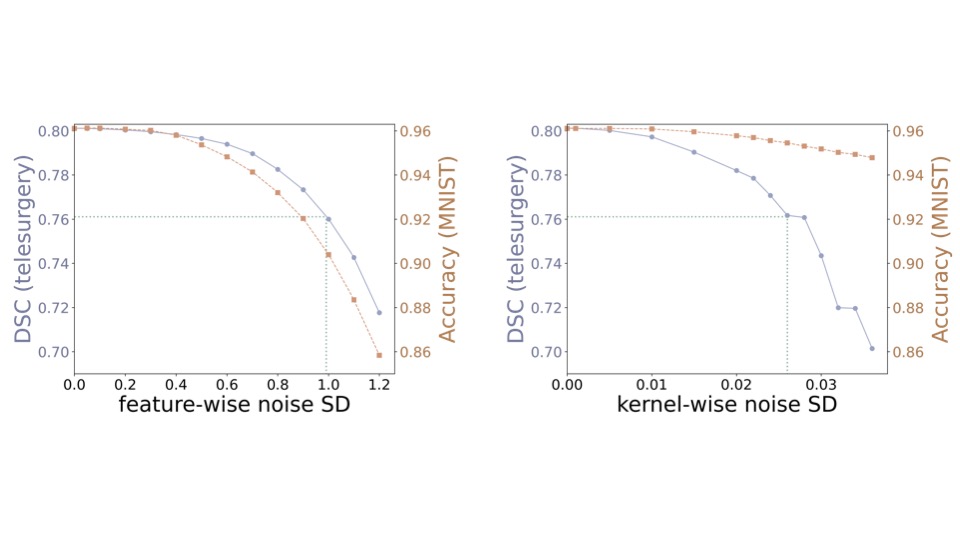}
    \vspace{-0.5cm}
    \caption{
    \textbf{Noise robustness comparison between a 161-layer vascular segmentation model and an MNIST classifier, highlighting the critical importance of kernel precision and stability.} 
    Model performance was quantified using the Dice similarity coefficient for vascular segmentation and the classification accuracy for MNIST under feature noise (left) and kernel noise (right). Both models exhibit tolerance to feature noise within their respective tasks; however, performance markedly deteriorates under even modest kernel perturbations. The degradation due to kernel noise is particularly pronounced for the deeper surgical segmentation model, as structural errors progressively accumulate and intensify across layers. To maintain at least 95\% of baseline performance (Dice = 0.761), kernel noise should remain below $\sigma$ = 0.026 (where $\sigma$ denotes the standard deviation of additive Gaussian noise), approximately 38× lower than the corresponding threshold for feature noise ($\sigma$ = 1.0). These results underscore the stringent requirement for kernel precision in scalable optical deployments of deep neural networks. 
    }
    \label{fig:figure5}
\end{figure}

To evaluate the noise resilience of our \ac{OCiC} system, we assessed its robustness under controlled noise injection. Noise arises primarily from two sources: distortions in the convolutional optical filter and \ac{AWGN} arising throughout the optical and electronic subsystems. The former perturbs convolution kernels, whereas the latter corrupts feature maps at each layer. The results are shown in Extended Data Fig.~\ref{fig:figure5}.

For kernel noise, given a 3×3 kernel $\mathbf{K}$, the perturbed kernel is defined as 
\begin{equation}
\begin{aligned}
\mathbf{K}_n = \mathbf{K} + \mathbf{N}
\end{aligned}
\end{equation}
where $\mathbf{N} \sim \mathcal{N}(0, \sigma^2 I)$, $\sigma$ denotes the kernel noise level, and $\mathbf{N} \in \mathbb{R}^{3 \times 3}$.

\vspace{0.5cm}
For feature-map noise, given a convolutional feature map F of shape $[b, c, h, w]$, the perturbed feature map is 
\begin{equation}
\begin{aligned}
\mathbf{F}_n = \mathbf{F} + \hat{\mathbf{N}}
\end{aligned}
\end{equation}
where $b$ is the batch size, $c$ the number of channels, and $h$ and $w$ are the spatial dimensions. Here $\hat{\mathbf{N}} \sim \mathcal{N}(0, \hat{\sigma}^2 I)$, $\hat{\sigma}$ denotes the feature noise level and  $\hat{\mathbf{N}} \in \mathbb{R}^{b \times c \times h \times w}$.

\vspace{0.5cm}
Having defined the two noise types, we next evaluated their impact. As shown in Extended Data Fig.~\ref{fig:figure5}, both models exhibit similar resilience to feature noise, with the deep segmentation network maintaining at least 95\% of its original accuracy until the noise standard deviation reaches 1. For the MNIST classifier, performance drops below 95\% of baseline when the feature noise standard deviation approaches 0.95. In contrast, kernel noise has a substantially greater impact across network depths, with the effect amplified in the 161-layer surgical segmentation model compared to the 1-layer MNIST classifier. To preserve 95\% accuracy in the segmentation model, the kernel noise standard deviation must remain below 0.026, corresponding to only a 0.7\% performance degradation in the MNIST classifier. This discrepancy arises because feature noise is largely independent across layers, whereas kernel noise directly perturbs the convolution process and accumulates progressively throughout the network. Such stringent noise tolerance has seldom been addressed in prior optical computing studies, highlighting the importance of stable and precise kernel assignment for scalable optical computing.

\newpage
\bibliographystylemethods{unsrt}
\bibliographymethods{ref_methods}

\section*{Data Availability}
The authors declare that the data underlying this study are provided within the paper and its accompanying supplementary materials. 

The MNIST handwritten digits dataset is available at \url{https://git-disl.github.io/GTDLBench/datasets/mnist_datasets/}. 

The dataset used for segmentation of coronary arteries in x-ray angiogram is available via web page (\url{http://personal.cimat.mx:8181/~ivan.cruz/DB_Angiograms.html}). 

The supplementary video of the comb stability measurement for the NDFF deployment can be found via (\url{https://drive.google.com/file/d/1PreiWbMKAI9J2nq43XG8fZdBkIGRxHoM/view?usp=drive_link}).

\section*{Code Availability}

The code for this paper can be available from the authors on request.

\section*{Acknowledgements}
This work was supported by the National Dark Fibre Facility (NDFF) funded by the Engineering and Physical Sciences Research Council [grant number EP/S028854/1] in the United Kingdom.
J.-Q. Z. acknowledges the Kennedy Trust Prize Studentship (AZT00050-AZ04) and the Chinese Academy of Medical Sciences (CAMS) Innovation Fund for Medical Science (CIFMS), China (grant number: 2024-I2M-2-001-1). 

\section*{Author contributions statement}
R.Y. and J.Z. conceived the OCiC architecture. R.Y. led the photonic experiments, including design, simulation, construction and testing, assisted by Y.L. J.H. developed the AI network and performed training, with support from J.Z. and Z.W.; J.Z. also conducted data curation. Q.R. and M.T. optimised the optical system design. J.C. and D.Z. contributed clinical and robotics expertise. J.H. and J.Z. carried out data analysis and visualisation. R.Y., Y.L., Y.Y., J.E.W. and X.L. prepared and deployed the NDFF experiments. R.Y., J.H., J.Z., Y.Y. and J.E.W. prepared figures and drafted the manuscript. All authors reviewed and edited the manuscript. C.L., H.L. and C.M. supervised the project.
\end{document}